\documentstyle[epsf,prd,twocolumn,aps,floats]{revtex} 
\tightenlines
\begin{document}

\draft
\topmargin = -0.6cm
\topmargin = -2.0cm
\overfullrule 0pt
\twocolumn[\hsize\textwidth\columnwidth\hsize\csname
@twocolumnfalse\endcsname

\title{
\vglue -0.5cm
\vglue 0.5cm
Neutrinoless double beta decay with and without Majoron-like boson\\ emission
 in a 3-3-1 model} 

\author{ J. C. Montero$^1$~\thanks{E-mail address: montero@ift.unesp.br}, 
C. A. de S. Pires$^2$~\thanks{E-mail address: cpires@ift.unesp.br},
V. Pleitez$^1$~\thanks{E-mail:vicente@ift.unesp.br}  
}  
\address{
$^1$ Instituto de F\'\i sica Te\'orica\\
Universidade Estadual Paulista\\
Rua Pamplona 145\\ 
01405-900--S\~ao Paulo, SP\\Brazil\\
$^2$ Department of Physics, University of Maryland\\
    College Park, MD 20742-4111, USA}

\date{\today}
\maketitle
\vspace{.5cm}

\hfuzz=25pt
\begin{abstract}
We consider the contributions to the neutrinoless double beta decays
in a $SU(3)_L\otimes U(1)_N$ electroweak model. We show that for a 
range of the parameters in the model there are 
diagrams involving vector-vector-scalar and trilinear scalar couplings
which can be potentially as contributing as the light massive Majorana
neutrino exchange one. 
We use these contributions to obtain constraints upon some mass
scales of the model, like the masses of the new charged vector and scalar 
bosons. We also consider briefly the decay in which
besides the two electrons a Majoron-like boson is emitted. 

\end{abstract}
\pacs{PACS numbers: 23.40.-s; 12.60. Cn; 14.60.St }



\vskip2pc]

\section{Introduction}
\label{sec:intro}

The issue of neutrino masses continues to be a golden plate in elementary
particle physics. Although data coming from solar~\cite{expsolar}, 
atmospheric~\cite{expatm}, and the accelerator LSND~\cite{lsnd} neutrino
experiments strongly suggest that neutrinos must be 
massive particles, direct measurements did not obtain any positive
result~\cite{pdg}. 

It is a very well known fact that if neutrinos are massive Majorana particles 
it should exist the neutrinoless double beta $(\beta\beta)_{0\nu}$ 
decay~\cite{rosen,rev}.
If the neutrino mass is the main effect that triggers this decay, the 
decay lifetime is proportional to (for the case of light neutrinos)
\begin{equation}
\langle M_\nu\rangle=\sum_iU^2_{ei}m_{\nu_i},
\label{lifetime}
\end{equation}
where $U_{ei};\,i=1,2,3$ denote the elements of a mixing matrix that relates 
symmetry $\nu_\alpha;\,\alpha=e,\mu,\tau $ and mass eigenstates 
$\nu_i$ through the relation $\nu_\alpha=\sum_iU_{\alpha i}\nu_i$; 
and $m_{\nu_i}$ are the neutrino masses. Experimentally a half life limit
$T^{0\nu}_{1/2}>1.8\times10^{25}$ yr implies~\cite{exp}
\begin{equation}
\langle M_\nu\rangle<0.2\;\mbox{eV}.
\label{ul}
\end{equation} 

The important point is that the $(\beta\beta)_{0\nu}$ decay probes the physics
beyond the standard model. In particular the observation of this decay would
be an evidence for a massive Majorana neutrino although it could 
say nothing about the value of the mass. This is because
although right-handed currents and/or scalar-bosons may affect the decay rate, 
it has been shown that whatever the mechanism of this decay is, a nonvanishing 
neutrino mass is required for the decay to take place~\cite{sv1}. 
However, this does not mean that the neutrino mass is necessarily the 
main factor triggering this decay. 
In some models the  $(\beta\beta)_{0\nu}$ decay 
can proceed with arbitrary small neutrino mass via scalar boson 
exchange~\cite{pt}.
The mechanism involving a trilinear interaction of the scalar bosons was 
proposed in Ref.~\cite{mv} in the context of a model
with $SU(2)\otimes U(1)$ symmetry with doublets and a triplet of scalar
bosons.
However, since in this type of models there is no large mass scale, it was
shown in Refs.~\cite{hw} that the contribution of the trilinear interactions 
are in fact negligible. 
In general, in models with that symmetry,  a fine tuning is needed
if we want that the trilinear terms give important contributions to
the $(\beta\beta)_{0\nu}$ decay~\cite{sv1,ep}. 

Here we will show that in a model with gauge symmetry $SU(3)_c\otimes
SU(3)_L\otimes U(1)_N$ (3-3-1 model by short)~\cite{331}, which has a rich 
Higgs bosons sector as in the multi-Higgs extensions
of the standard model, there are new contributions to the 
$(\beta\beta)_{0\nu}$ decay. However, unlike the latest sort of models, 
a fine tuning of the parameters of
the 3-3-1 model is not necessary since some trilinear couplings, 
which have mass dimension, could imply an enhancement of the respective 
amplitudes (See. Sec.~\ref{sec:bb0n}).

We will use the following strategy: First, we consider the several
new contributions to the $(\beta\beta)_{0\nu}$ decay introduced by the
3-3-1 model. Next, once this decay has not experimentally been seen, we will
consider the usual standard model amplitude (that would arise with
massive Majorana neutrinos) as the reference one and make the assumption
that all the new amplitudes are at most as contributing as this 
one. Hence, we can obtain constraints on some typical mass scale 3-3-1 
parameters. The new contributions to the $(\beta\beta)_{0\nu}$ decay are
of the short range type~\cite{srange}. 
Since the respective matrix elements are different
from those of the long range contributions (the exchange of a 
light-Majorana neutrino) our results should be considered only as
an indication of the possible large contributions to this decay in the
context of the 3-3-1 model.   

The outline of the paper is the following. In Sec.~\ref{sec:model} we introduce
the interactions which are relevant to the present study. The model with
$\langle \sigma^0_1\rangle\not=0$, which is some 
cases it has a Majoron-like Goldstone boson, is also discussed. 
In Sec.~\ref{sec:bb0n} we consider the more important contributions
to the $(\beta\beta)_{0\nu}$ decay and the constraints upon some 
masses of the model. In Sec.~\ref{sec:me} we show that if we add a neutral
scalar singlet to the minimal model a Majoron-like Goldstone boson is consistent
with the $Z^0$ invisible width and we also
discuss briefly the Majoron emission process $(\beta\beta)_{0\nu M}$
comparing the relative strength of two amplitudes.
Our conclusions remain in the last section.

\section{The model}
\label{sec:model}

Here we will consider the 3-3-1 model with the leptons belonging to
triplets $(\nu_l\,l\,l^c)^T;\,l=e,\mu,\tau$ and in which a sextet
of scalar bosons
\begin{equation}
S=\left( 
\begin{array}{ccc}
\sigma _1^0 & \frac{h_2^{-}}{\sqrt2} & \frac{h_1^{+}}{\sqrt2} \\ 
\frac{h_2^{-}}{\sqrt2} & H_1^{--} & \frac{\sigma _2^0}{\sqrt2} \\ 
\frac{h_1^{+}}{\sqrt2} & \frac{\sigma _2^0}{\sqrt2} & H_2^{++}
\end{array}
\right) \sim \left( {\bf 6},{\bf 0}\right).
\label{s}
\end{equation}
is necessary to give to the charged leptons a mass if 
$\langle\sigma^0_2\rangle\equiv v_{\sigma_2}\not=0$~\cite{331}.

Most of the phenomenological studies of the model has been done by 
considering $\langle \sigma^0_1\rangle\equiv v_{\sigma_1}=0$. The 
case when $\langle \sigma^0_1\rangle\not=0$ it was considered in
Ref.~\cite{ma1}, where the other scalar multiplets are explicitly given. 
The main difference in the latter case with respect to the
former one is that there is a mixing between the vector bosons $W^+$ and 
$V^+$:

\begin{equation}
\left(\begin{array}{cc}
 W^+_\mu & V^+_\mu
\end{array}
\right)
\left( 
\begin{array}{cc}
M^2_W & \delta \\
\delta & M^2_V
\end{array}
\right)
\left(
\begin{array}{c}
W^-_\mu\\ V^-_\mu
\end{array}
\right),
\label{mm}
\end{equation}
where $\delta=(g^2/2)(2v_{\sigma_1}v_{\sigma_2})$, $M^2_{W}$ and 
$M^2_V$ are the mass eigenvalues
when $\delta=0$; if $\delta\not=0$ ({\it i.e.} when $v_{\sigma_1}\not=0)$ 
the mass of the physical fields are now 
\begin{equation}
2M_{1,2}=(M^2_W+M^2_V) 
\pm\left[(M^2_W-M^2_V)^2+4\delta^2\right]^{1/2}
\label{massnow}
\end{equation}
and we have defined
$M^2_W=(g^2/2)(v^2_\eta+v^2_\rho+v^2_{\sigma_2}+v^2_{\sigma_1})$,
$M^2_V=(g^2/2)(v^2_\eta+v^2_\chi+v^2_{\sigma_2}+v^2_{\sigma_1})$ and
$g$ is the $SU(3)_L$ coupling constant which is numerically equal to the
coupling of  
$SU(2)_L$ i.e., $g^2=8M^2_WG_F/\sqrt2$. We have denoted by $v_\eta$, $v_\rho$
and $v_\chi$ the vacuum expectation values of the neutral components of
the triplets.
Notice that $M_1\to M_W$ and $M_2\to M_V$ when $\delta\to 0$.  
The vector bosons $W^+_{\mu}$ and $V^+_{\mu}$ are related to the new mass 
eigenstates $W^+_{1\mu}$ and $W^+_{2\mu}$ as
\begin{equation}
\left(
\begin{array}{c}
W^+ \\ V^+
\end{array}
\right)=\left(
\begin{array}{cc}
c_\theta & s_\theta\\
-s_\theta & c_\theta
\end{array}
\right)
\left(
\begin{array}{c}
W^+_{1} \\ W^+_{2}
\end{array}
\right)
\label{me}
\end{equation}
with $\tan2\theta=-2\delta/(M^2_W-M^2_V)$.
We can obtain an upper bound on $\delta$ by assuming that the main contribution
to the $M^2_W$ mass is given by $v_{\sigma_2}\approx 246$ GeV and
that $v_{\sigma_1}$ has its maximum value 3.89 GeV allowed by the value of the
$\rho$-parameter~\cite{comment}. In fact if $v_{\sigma_2}$ were the main 
contribution to the $M_W$ mass we would have $\delta/M^2_W\approx 
(2v_{\sigma_1}/v_{\sigma_2})<0.032$.
The constraint on the mixing angle $\theta$ is:
\begin{equation}
0\leq s^2_\theta=\frac{1}{2}\left(1-\frac{M^2_V-M^2_W}
{[4\delta^2+(M^2_V-M^2_W)^2]^{1/2}}
 \right)<\frac{1}{2}.
\label{sin}
\end{equation}

Some illustrative values for $s_\theta$ are obtained 
by using typical values for the parameters. For instance, 
for $v_{\sigma_1}=3.89$ GeV; $v_{\sigma_2}=10$ GeV, $M_W=80.41$ GeV and 
$M_V=100\,(300)$ GeV we get $s^2_\theta=1.9\times10^{-5}(3.4\times10^{-8})$;
or if $M_V=100 $ GeV and if $v_{\sigma_2}$ has its maximal value 
$v_{\sigma_2}=246$ GeV we have $s^2_\theta=1.1\times10^{-2}$. 
We see that only for values of $M_V\approx M_W$ 
the $s^2_\theta$ is almost 0.5 but this light vector boson may be not 
phenomenologically safe.  
However if $v_{\sigma_1}$ is of the same order of magnitude of the neutrino 
mass smaller values for the mixing angle are obtained. 
Hence, it may be no relevant for the collider physics and low energy 
processes like the $(\beta\beta)_{0\nu}$ decay  
at all and in practice $W^+_1\approx W^+$, $W^+_2\approx V^+$; but this 
could not be the case in astrophysical processes~\cite{ma1}.  

Next, we consider the several interactions that are present in this model.
The scalar-quark interactions are
\begin{eqnarray}
-{\cal L}_Y^{u-d}&\!=\!&\frac{\sqrt2}{\vert v_\rho\vert}\,
\overline{D}_L V^\dagger_{\rm CKM} M^u U_R\, 
\rho ^{-}\!\!\!+\!\!
\frac{\sqrt{2}}{\vert v_\eta\vert}\overline{U}_L
V_{\rm CKM} M^d{ D}_R
\, \eta _1^{+}
\nonumber \\ &&\mbox{}
+\overline{D}_L\left( V_L^d\right) ^T\triangle \text{ }V_L^u
M^u U_R  \left[\frac{\sqrt{2}}{\vert v_\eta\vert}\, \eta_1^{-}-
\frac{\sqrt{2}}{\vert v_\rho\vert }\, \rho ^{ -}\right]
\nonumber \\ &&\mbox{}+ 
\overline{U}_L\left( V_L^u\right) ^T\triangle \text{ }V_L^d
M^d D_R\left[ \frac{\sqrt{2}}{\vert v_\rho\vert }\, 
\rho ^{+}- \frac{\sqrt{2}}{\vert v_\eta\vert}\,\eta _1^{ +}\right] 
\nonumber \\ &+& H.c. , 
\label{yuka}
\end{eqnarray}
with 
\begin{equation}
\Delta \equiv \left(
\begin{array}{ccc}
0 & 0 & 0 \\ 
0 & 0 & 0 \\ 
0 & 0 & 1
\end{array}
\right),
\label{delta}
\end{equation}
and $V^{u,d}_L$ are unitary mixing matrices, and $M^{u,d}$ are the diagonal
mass matrices of the $u$-like and $d$-like quark sectors, and $V_{\rm CKM}$
denotes the usual mixing matrix of Cabibbo-Kobayashi-Maskawa.

The Yukawa interactions in the lepton sector are
\begin{eqnarray}
-{\cal L}^l_Y&=&\frac{1}{\sqrt2}\,\overline{\nu_L}\,{\cal K}_{1}
l_RH^+_1+
\frac{1}{\sqrt2}\,\bar{l}_L\,{\cal K}_{2} \nu^c_RH^-_2
\nonumber \\ &+&
\frac{1}{2}\,\overline{l_L}\,{\cal K}_{3}(l_L)^cH^{--}_1 
+\frac{1}{2}\,\overline{(l^c)_L}\,{\cal K}_{4}l_RH^{++}_2
\nonumber \\ \mbox{}
&+&
 2\;\overline{\nu_L}\, {\cal K}^\prime_{1}l_R\eta^+_1
-2\;\overline{l_L}\,{\cal K}^\prime_{2}\nu^c_R\eta^-_2
+H.c.,
\label{yu331}
\end{eqnarray}
where ${\cal K}_{1}=E^{\nu\dagger}_LGE_R$, ${\cal K}_{2}= 
E^{\dagger}_LGE^{\nu*}_L$; ${\cal K}_{3}=E^{\dagger}_LGE_L$,
${\cal K}_{4}=E^T_RGE_R$; ${\cal K}^\prime_{1}=
E^{\nu\dagger}_L G^\prime E_R$, ${\cal K}^\prime_{2}=
E^{\dagger}_LG^\prime E^{\nu*}_L$;
$G$ and $G'$ are symmetric and antisymmetric (they can be complex) matrices,
respectively. $E_{R},E_L,E^\nu_L$ are the right- and 
left-handed mixing unitary matrices in the lepton sector relating symmetry 
eigenstates (primed fields) with mass eigenstates 
(unprimed fields)~\cite{liung}:
\begin{equation}
l'_R=E_Rl_R,\quad l'_L=E_Ll_L,\quad  \nu'_L=E^\nu_L\nu_L,
\label{sme}
\end{equation}

Some of the couplings in Eq.~(\ref{yu331}) do not depend on  the 
charged lepton masses and since all matrices in Eq.~(\ref{yu331}) are
not unitary, the model breaks the lepton universality but it can be
shown that, for the massless neutrino case no strong 
constraints arise from exotic muon and tau decays~\cite{fcnit}. 

In the scalar sector we have also mixing angles. In the singly charged sector we
have $\phi^-_i=\sum_lO_{ij}H^-_j$,   
where $\phi^-_i=\eta^-_1,\eta^-_2,\rho^-,\chi^-,h^-_1,h^-_2$ and
$H^-_j,\,j=1,...,6$ denotes the respective mass eigenstate field; 
similarly in the doubly charged sector
we have $\Phi^{--}_i=\sum_l{\cal O}_{ij}\Psi^{--}_j$, with 
$\Psi^{--}_i=\rho^{--},\chi^{--}, H^{--}_1,H^{--}_2$ and 
$\Psi^{--}_j,\,j=1,..,4$ the respective mass eigenstate fields. However,
in the following we will use $H^-$ and $H^{--}$ as typical mass eigenstates
of the respective charged fields and omit the scalar mixing parameters. 

We recall that the model conserves the ${\cal F}=L+B$ quantum number; 
$L$ is the total lepton number and $B$ is the baryon number. 
The assignments are:
\begin{eqnarray}
{\cal F}(U^{--})&=&{\cal F}(V^{-}) =
{\cal F}(\rho^{--}) = {\cal F}(\chi^{--})={\cal F}(\eta^-_2)
\nonumber \\&=&{\cal F}(\chi^{-})
= {\cal F}(\sigma_1^0)={\cal F}(h^-_2)={\cal F}(H^{--}_1)
\nonumber \\ &=&{\cal F}(H^{--}_2)
=2,
\label{efe}
\end{eqnarray}
and all other scalar fields with ${\cal F}=0$. 

The charged currents coupled to the vector bosons are given by 
 \begin{eqnarray}
{\cal L}^{CC}&=&-\frac{g}{\sqrt{2}}\left(\overline{U}_L\gamma^u
V_{\rm CKM}D_LW^+_{\mu}
-\bar\nu_L\gamma^\mu V_W l_LW^+_{\mu}\right. 
\nonumber \\ &+& 
\left.\overline{l^c_L}\gamma^\mu V_UV^\dagger_W  \nu_L V^+_{\mu}-
\overline{l^c_L} \gamma^\mu V_Ul_L U^{++}_\mu\right)+H.c,
\label{v331}
\end{eqnarray}
with the mixing matrices defined as $V_{\rm CKM}=
\left( V_L^u\right)^\dagger V_L^d$ in the quark sector; and 
$V_W={E^\nu_L}^\dagger E_L$, $V_U=E^T_RE^\nu_L$ in the leptonic sectors.  

We have the trilinear interactions involving one vector- and two scalar-bosons 
which are of the form (up to a $ig/\sqrt2$ factor):
\begin{eqnarray}
{\cal L}^{V2S}&=&\partial^\mu \chi^+ \chi^{--}W^+_\mu
+\chi^-\partial^\mu \chi^{++}W^-_\mu+
\partial^\mu h^+_2 H^{--}_1 W^+_\mu
\nonumber \\ & +& \partial^\mu H^{++}_1 h^-_2 W^-_\mu
+\partial^\mu\rho^-\rho^{++}V^-_\mu+ \partial^\mu\rho^{--}\rho^+V^+_\mu
\nonumber \\ &+&
h^+_1\partial^\mu H^{--}_2 V^+_\mu+\partial^\mu h^-_1 H^{++}_2 V^-_\mu
+\left(\eta^+_2\partial^\mu\eta^+_1\right.\nonumber \\ &+&
\left.\eta^+_1\partial^\mu\eta^+_2 +h^+_2\partial^\mu h^+_1+
h^+_1\partial^\mu h^+_2\right) U^{--}_\mu+H.c.\,. 
\label{eev}
\end{eqnarray}

There are also trilinear interactions involving two vector- and one 
scalar-bosons (the $\chi^-$ scalar couples to ordinary and exotic quarks and for
this reason it is not of our concern here). They are given by (up to a $g^2/2$ 
factor)
\begin{eqnarray}
{\cal L}^{2VS}&=&\frac{v_{\sigma_1}}{2}\,H^{++}_1
W^{-}\cdot W^-+  \frac{v_{\sigma_1}}{2}\,H^{++}_2V^{-}\cdot V^-
\nonumber \\ &+&
\left( v_\rho \rho^{++}+v_\chi\; \chi^{++}+
\frac{v_{\sigma_2}}{2}(H^{++}_1+H^{++}_2)\right)
\nonumber \\ &&\mbox{} W^{-}\cdot V^-.
\label{vve}
\end{eqnarray}
Notice that there is a coupling which is proportional to $v_\chi$ and
hence it will be the dominant one. 

Next, we write down the trilinear interactions among three vector bosons
\begin{eqnarray}
{\cal L}^{3V}&=&i\frac{g}{\sqrt2}\left(W^+_{\mu\nu}V^{\nu+}U^{\nu--}
+V^+_{\mu\nu}V^{\nu+}U^{\mu--}\right.\nonumber \\ &+&
\left.W^{\nu+}V^{\mu+}U^{--}_{\mu\nu}\right)
\label{3v}
\end{eqnarray}
where $W_{\mu\nu}=\partial_\mu W_\nu-\partial_\nu W_\mu$ and so on. 

Finally,
we have trilinear interactions among scalar-bosons only
\begin{equation}
{\cal L}^{3S}=\frac{f_1}{2}\,\epsilon^{ijk}\eta_i\rho_j\chi_k+
\frac{f_2}{2}\,\chi^TS^\dagger\rho
+H.c. .
\label{3e}
\end{equation}
The couplings $f_{1,2}$ have dimension of mass but are both of them 
arbitrary parameters (see next section). Other terms like the trilinears 
$f_3\eta^T S^\dagger\eta$ and $f_4 \epsilon SSS$ and the quartic interactions
$\epsilon \chi(S\eta^*)\rho$, $\chi^\dagger\eta\rho^\dagger\eta$ and 
$\epsilon\chi\rho SS$ violate the conservation of ${\cal F}$. 

However, as we will show in Sec.~\ref{sec:me}, when discussing the Majoron
emission, the model must be modified by adding a scalar singlet in order to be
consistent with the LEP data.

\section{The neutrinoless double beta decay}
\label{sec:bb0n}

\begin{figure}[ht]
\centering\leavevmode
\epsfxsize=200pt
\epsfbox{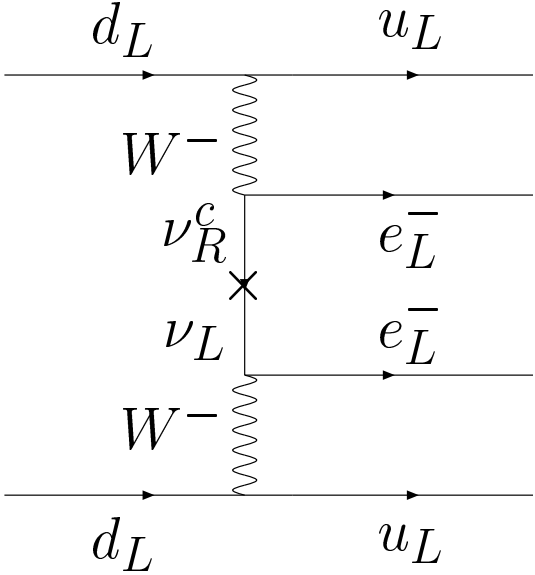}
\vglue -0.01cm
\caption{$(\beta\beta)_{0\nu}$ decay via light massive
Majorana neutrinos.}
\label{fig1}
\end{figure}

\begin{figure}[ht]
\centering\leavevmode
\epsfxsize=100pt
\epsfbox{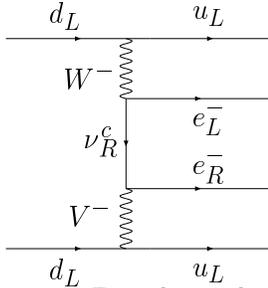}
\vglue -0.01cm
\caption{The same as in Fig.~1 but with one bilepton vector boson $V^-$ 
replaced by a vector boson $W^-$.}
\label{fig2}
\end{figure}

\begin{figure}[ht]
\centering\leavevmode
\epsfxsize=100pt
\epsfbox{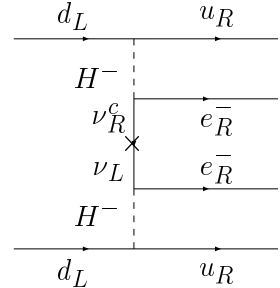}
\vglue -0.01cm
\caption{Charged scalars contribution to the $(\beta\beta)_{0\nu}$ decay.   }
\label{fig3}
\end{figure}

\begin{figure}[ht]
\centering\leavevmode
\epsfxsize=100pt
\epsfbox{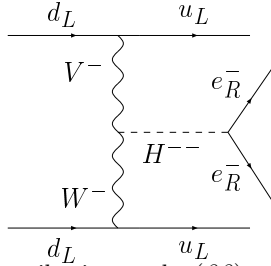}
\vglue -0.01cm
\caption{New contribution to the $(\beta\beta)_{0\nu}$ decay in the 3-3-1 
model.   }
\label{fig4}
\end{figure}

\begin{figure}[ht]
\centering\leavevmode
\epsfxsize=100pt
\epsfbox{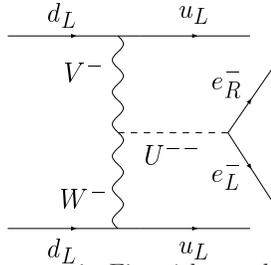}
\vglue -0.01cm
\caption{The same as in Fig. 4 but only involving vector bosons.   }
\label{fig5}
\end{figure}

\begin{figure}[ht]
\centering\leavevmode
\epsfxsize=100pt
\epsfbox{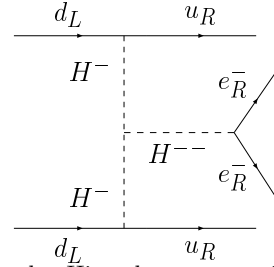}
\vglue -0.01cm
\caption{Pure scalar Higgs bosons contribution the $(\beta\beta)_{0\nu}$ 
decay in the 3-3-1 model.  }
\label{fig6}
\end{figure}

Some of the more relevant diagrams of the $(\beta\beta)_{0\nu}$ decay
in the present model are shown in Figs.~\ref{fig1}-\ref{fig6}.
Our goal is to analyze the order of magnitude
of each diagram and to obtain constraints on some mass scales of the model. 
We will consider the diagram in 
Fig.~\ref{fig1} as the reference one, i.e., it is the diagram that already 
exist in the standard 
model framework with massive Majorana neutrinos and which is
parameterized by two effective four-fermion interactions. The other 
contributions will be considered as been at most equally important than
the standard one.

\vglue 0.01cm

The strength of the diagram in Fig.~\ref{fig1} is given by
\begin{equation}
A(1)\propto\frac{g^4 \langle M_\nu\rangle}{M^4_W\langle p^2\rangle}\,
c^4_\theta
=\frac{32G^2_F\langle M_\nu\rangle}{\langle p^2\rangle}\,c^4_\theta,
\label{a1}
\end{equation}
where $\langle M_\nu\rangle$ is the effective 
mass defined in Eq.~(\ref{lifetime}) and $\langle p^2\rangle$ is the average 
of the four-momentum transfer squared, which is of the order of
$(100\,\mbox{MeV})^2$. Below we will use a small $\delta$ so that $M_1\to M_W$
and $M_2\to M_V$. 

In Eq.~(\ref{a1}) and hereafter we will omit for simplicity the mixing 
parameters. 
Only in the vertices we will take care about the mixing between $W$ 
and $V$ defined in Eq.~(\ref{me}) but in the propagator we will use the 
masses of $W$ and $V$.

\vglue 0.01cm

Next, let us consider the diagram in Fig.~\ref{fig2} which has the strength
given by 
\begin{equation}
A(2)\propto 
32G^2_F\,\left(\frac{M_W}{M_V}\right)^2
\frac{c^3_\theta s_\theta}{\sqrt{\langle p^2\rangle}},
\label{a2}
\end{equation}
and we have the ratio
\begin{equation}
\frac{A(2)}{A(1)}=\left( \frac{M_W}{M_V}\right)^2
\frac{\sqrt{\langle p^2\rangle}}{\langle M_\nu\rangle}\tan\theta,
\label{a2a1}
\end{equation}
and if $A(2)/A(1)<1$ we have that 
\begin{equation}
M_V>2.2\times 10^4 M_W\sqrt{\tan_\theta}=1.79\times10^6\sqrt{\tan_\theta}\,
{\rm GeV}.
\label{c1}
\end{equation}

We recall that a lower limit of 440 GeV is obtained for $M_V$ from the muon 
decays but when only the bilepton contributions to those decays are
considered~\cite{joshi}. 
However, in the minimal 3-3-1 model the scalar-boson contributions cannot 
be negligible since some of the charged scalar-bosons can be lighter than the 
vector bilepton boson $V^-$. 
Hence, a lighter vector boson $V$ may still be possible but this 
subject deserves a more detailed study of the muon decay considering both
vector and scalar contributions. A contribution similar to that in 
Fig.~\ref{fig1} but with two $V^-$ bosons instead of two $W^-$ bosons
may be not negligible but it does not constraint the mass $M_V$ 
as much as those in Eq.~(\ref{c1}) since the condition that its ratio
to the $A(1)$ amplitude be less than one gives the condition 
$M_V>M_W\sqrt{\tan\theta}$.

All the Lagrangian interactions in Eqs.~(\ref{yuka}), 
(\ref{yu331}), (\ref{v331}), (\ref{eev}), (\ref{vve}) and (\ref{3v}) 
are written in terms of symmetry eigenstates.
We have assumed Yukawa couplings of the order of unity. As we are not 
considering the mixing among the scalar fields 
our constraints
are valid only for the main component of the symmetry eigenstate scalar 
fields. It means that $H^-$ and $H^{--}$ denote the dominant mass eigenstates
of the singly and doubly charged scalar fields, respectively.

The amplitude of the diagram in Fig.~\ref{fig3} is
\begin{equation}
A(3)\propto
\frac{\langle M_\nu\rangle}{\langle p^2\rangle M^4_{H^{-}}}.
\label{a3}
\end{equation}

\vglue 0.01cm

The scalar contribution in Fig.~\ref{fig3} can be as 
important as the standard one in Fig.~\ref{fig1}. We have
\begin{equation}
\frac{A(3)}{A(1)}=
\frac{1}{32G^2_FM^4_{H^{-}}c^4_\theta},
\label{a3a1}
\end{equation}
and assuming that $A(3)/A(1)<1$ and $c_\theta=1$ we get
\begin{equation}
M_{H^{-}}>124\,\mbox{GeV}.
\label{c2}
\end{equation}

From Eq.~(\ref{vve}) we see that the contribution $v_\chi\chi^{++}
W^-_\mu V^{\mu-}$ is the dominant one in diagrams like that in 
Fig.~\ref{fig4}. As we said before we will
omit the mixing angles, i.e., assuming $\chi^{--}\approx H^{--}$.
Hence we have
\begin{equation}
A(4)\propto\frac{v_\chi}{M^2_WM^2_VM^2_{H^{--}}}c^2_\theta s^2_\theta.
\label{a4}
\end{equation}

\vglue 0.01cm

Next, we note that
\begin{eqnarray}
\frac{A(4)}{A(1)}&\propto& \frac{v_\chi}{\langle M_\nu\rangle}\,
\frac{\langle p^2\rangle}
{32G^2_FM^2_WM^2_VM^2_{H^{--}}}\tan^2\theta,\nonumber \\&&\mbox{}
 \approx 5.33 \times10^{15}\,
\tan^2\theta\,\frac{(1{\rm GeV})^4}{M^2_VM^2_{H^{--}}},
\label{a4a1}
\end{eqnarray}
where we used $\langle M_\nu\rangle=0.2$ 
eV~\cite{pdg}, $v_\chi=3$ TeV and   
$\langle p^2\rangle=(100\,\mbox{MeV})^2$. 
If $A(4)/A(1)<1$ it implies 
\begin{equation}
M_V>7.3 \times 10^{7}\,\tan\theta\,\frac{(1{\rm GeV})^2}{M_{H^{--}}}.
\label{c3}
\end{equation} 
 
Similar analysis arises by considering Fig.~\ref{fig5}, however it is less 
enhanced than the contribution of Fig.~\ref{fig4} because instead of $v_\chi$
it appears the momentum of one of the vector bosons, 
$p\sim\sqrt{\langle p^2\rangle}$. 

\vglue 0.01cm

More interesting are the contributions involving trilinear scalar interactions 
given in Eq.~(\ref{3e}) like that of the diagram in Fig.~\ref{fig6}. 
We have in this case
\begin{equation}
A(6)\propto \frac{f}{M^4_{H^-}M^2_{H^{--}}},
\label{a5}
\end{equation}
where $M_{H^{-}}$ represents a typical mass of the singly charged scalar 
bosons, say 124 GeV; 
$M_{H^{--}}$ is the mass of the doubly charged scalar boson and $f$ is 
the trilinear coupling $f_1$ or $f_2$ in Eq.~(\ref{3e}) with dimension of 
mass. The ratio of these amplitudes is: 
\begin{eqnarray}
\frac{A(6)}{A(1)}&\propto& \frac{f \langle p^2\rangle}
{32G^2_FM^4_{H^-}M^2_{H^{--}}\langle M_\nu\rangle c^4_\theta}\nonumber \\
&\approx& \left(\frac{f/\mbox{GeV}}{M^2_{H^{--}}/\mbox{GeV}^2}\right)
\cdot\frac{4.8\times 10^{7}}{c^4_\theta},
\label{a6a1}
\end{eqnarray}

If $A(6)/A(1)<1$, and assuming $c_\theta=1$ and $M_{H^-}=124$ GeV, 
we obtain the constraint
\begin{equation}
\frac{f}{M^2_{H^{--}}}<2.1\times10^{-8}\,\mbox{GeV}^{-1}.
\label{c4} 
\end{equation}

For  arbitrary $U(1)_N$ charge for the scalar multiplets the symmetry
of the potential is $SU(3)_L\otimes [U(1)]^2$. If the triplet $\eta$ and the
sextet $S$ have both $N=0$, as it is the case for the present model, 
the trilinear couplings $f_{1,2}$ break
the extra $U(1)$ symmetry. We have verify that if both $f_{1,2}=0$
there is indeed a pseudo-Goldstone boson~\cite{sw}. 
It means that $f_{1,2}$ are
arbitrary parameters and in principle they can be small (say 1 GeV), or
large (say 1 TeV) mass scales. 
  
We see that if $f=1$ ($10^{-3}$) TeV then $M_{H^{--}}$ is greater or 
of the order of 300 (10) TeV. 
For this value for the mass of the doubly charged scalar field and $\theta$
small the constraint given in Eq.~(\ref{c1}) is stronger than that of 
Eq.~(\ref{c3}). For instance if $\tan\theta=10^{-8}$ we have from 
Eq.~(\ref{c1}) that $M_V\geq 179 $ GeV.

\vglue 0.01cm

There is also a diagram in which the doubly charged scalar field in 
Fig.~\ref{fig6} is substituted by a vector boson $U^{--}$. Although
the interactions in Eq.~(\ref{eev}) are proportional to $g$ they are 
also derivative and proportional to the momentum $p\sim\sqrt{\langle 
p^2\rangle}$; hence it is suppressed with respect to the diagram in 
Fig.~\ref{fig6}. 

\section{Majoron emission}
\label{sec:me}

If the ${\cal F}$ quantum number is spontaneously broken as in the present
model, it means that a Majoron-like bosons does exist. Since the scalar field
that is responsible for the breakdown of that continuous symmetry is
$\sigma^0_1$, and it belongs to a triplet of the subgroup $SU(2)\otimes U(1)$,
this Majoron-like Goldstone has similar couplings that the triplet Majoron model
of Ref.~\cite{gr}. 
It is well known that this sort of Majoron model has been ruled out by the LEP
data~\cite{concha}. Apparently, since the Higgs sector of the present model is
rather complicated having a neutral scalar singlet (under
$SU(2)\otimes U(1)$), $\chi^0$, it seems that the Majoron-like Goldstone in this
case it will be able to avoid the LEP constraints as claimed in
Refs.~\cite{ma1,ps}. However, we will show that this is not indeed the case. 

The mass matrices of the scalar and pseudoscalar in this model
have been given in Ref.~\cite{ma1}. Here we will only give the results of the
mass eigenvalues and the respective mixing matrix in the CP-even scalar
sector. 
The argument in Ref.~\cite{ma1} was the following. 
Let us begin with the relation $R^0_4=\sum_j{\cal O}^o_{4j}H^0_j$, 
where $H^0_j,\;j=1,...5,$ denotes the mass scalar eigenstates and $R_4$ the real
component of the scalar field $\sigma^0_1$ according to the general shifting of
the neutral scalar fields in the scalar potential of the form
$X^0_i\to(1/\sqrt2)(v_{X_i}+R_i+iI_i),\,i=1,2,3,4,5$ where
$X^0_i=\eta,\rho,\chi,\sigma_1,\sigma_2$  respectively. 
In this case if $H^0_1$ denotes the lightest scalar boson ($M_{H_1}<M_Z$, we are
assuming a mass spectrum where $M_{H_i}<M_{H_j}\,$ if $i<j$),  
the contribution to the
decay mode $Z^0\to H^0_1M^0$ is $\Gamma^Z_{H^0_1M^0}=
2\vert{\cal O}^o_{41}\vert^2\Gamma^Z_{\nu\bar\nu}$. 
Hence, if $\vert{\cal O}^o_{41}\vert<10^{-2}$ the model 
would be consistent with
the LEP data {\it i.e.}, now $\Gamma^Z_{H^0_1M^0}$ would be reduced to 
an acceptable level.

First of all recall that as shown in Ref.~\cite{ma1} the Majoron-like boson
decouples from the other pseudoscalar fields i.e., ${\rm Im}\sigma^0_1\equiv
I_4=M^0$. 
For instance, using the same values of the VEV and $f_{1,2}=-1$ TeV 
and with the dimensionless constant of the scalar potential given in
Ref.~\cite{ma1} with $\lambda_k= 0.1$ for $k=1,2,3,4,5,6,7,8,9,18,20$,
$\lambda_m=0.01$ for $m=13,14,16,17$, $\lambda_n=0.001$ for $n=10,11,12,19$ and
$\lambda_{15}=0.05$ we obtain the following masses in the scalar sector 
$(0.056,102, 1342, 3626, 4325)$ GeV and the mixing matrix (up to three decimal
places) 
\begin{equation}
O^o=\left( 
\begin{array}{ccccc}
0.0  & 0.081  & -0.010 & 0.996  & 0.021\\
0.0  & 0.995  & -0.029 & -0.082 & 0.039 \\
0.0  & -0.030 & -0.999 & -0.008 & 0.004 \\
1.0  & 0.0    & 0.0    & 0.0    & 0.0 \\
0.0  & 0.040  & -0.005 & 0.017  & -0.999\\
\end{array}
\right)
\label{ufa}
\end{equation} 
This pattern of mixing matrix remains the same for several values of the
parameters provided that $v_{\sigma_1}$ is a small VEV restricted to the
condition that it has to be smaller than 3.89 GeV~\cite{comment}. 
From Eq.~(\ref{ufa}) it can be seen that the scalar partner of the Majoron is
always mainly the lightest scalar i.e., $\vert{\cal O}^o_{41}\vert\sim1$ and it
would be always produced at LEP. We see that the Majoron in the minimal 3-3-1
model has been also ruled out by the LEP data. 

One possibility to recover consistency with the LEP data is to break explicitly
the ${\cal F}$ symmetry by adding trilinear terms like $f_3\eta^T
S\eta,f_4\chi^T S^\dagger \rho$ in the scalar potential, see Eq.~(\ref{3e}). In
this case there is no Majoron at all and although $v_{\sigma_1}$ still has a
small value, due to $f_3$ all scalars are heavy enough to not be produced at the
LEP energies~\cite{typeII,ema}. Of course, in this case there is no contribution
to Majoron emission in the neutrinoless double beta decay. However, our results
in Sec.~\ref{sec:bb0n} are still valid since they depend only on the small value
of $v_{\sigma_1}$.  

Another possibility which we will consider here is to modify the model by
introducing a scalar singlet $\Sigma^0$, which carries ${\cal F}=2$ (or $L=2$),
in the same way as considered in Ref.~\cite{sv,diaz} in the context of a 
$SU(2)\otimes U(1)$ model.
In this case we have to add the following terms to the scalar potential in
Ref.~\cite{ma1}
\begin{eqnarray}
V(X_i,\Sigma)&=&\mu^2_5\Sigma^2+\lambda_{21}
\Sigma^4+\sum_i\left[\lambda_{X_i}Tr\,(X^\dagger_i
X_i)\Sigma^2\right.\nonumber \\
&-&\left.\kappa\,\eta^TS^\dagger\eta\Sigma+H.c.\right]   
\label{sigma}
\end{eqnarray}
where $X_i$ denotes any triplet $\eta,\rho,\chi$ or the sextet $S$, and we will
denote $\lambda_{X_i}$ as $\lambda_{22,23,24,25}$ respectively and $\kappa>0$.
The neutral Higgs sector contains six CP-even scalars and three massive CP-odd 
pseudoscalar beside the massless CP-odd Majoron.
The neutral scalar singlet also gains a VEV, i.e.,
$\Sigma=(v_\Sigma+R_6+iI_6)/\sqrt2$, and the mass term is given by $M^2/2$,
where $M^2$ in the pseudoscalar sector in the basis $I_1,I_2,I_3,I_4,I_5,I_6$ is
given by (the constraints equation appear in the Appendix) 

\begin{eqnarray}
&&M_{11}= -\frac{\lambda_{16}}{2\sqrt{2}}\frac{v^2_\rho v_{\sigma_2}}
{v_\eta} +2\lambda_{17}v^2_{\sigma_2}
+ \frac{1}{2\sqrt{2}}(\lambda_{15}v_\chi v_{\sigma_2}\nonumber \\ &-&
f_1v_\rho) \frac{v_\chi}{v_\eta}+2\kappa v_{\sigma_1}v_\Sigma +
\frac{t_\eta}{v_\eta}, 
\nonumber \\&&
M_{22}=-\frac{1}{4}(\sqrt2 f_1 v_\eta +f_2v_{\sigma_2})\frac{v_\chi}{v_\rho}
+\frac{t_\rho}{v_\rho},\nonumber \\ \mbox{}&&
M_{33}=-\frac{1}{4}(\sqrt2 f_1 v_\eta+f_2
v_{\sigma_2})\frac{v_\rho}{v_\chi} +\frac{t_\chi}{v_\chi},
\nonumber \\ &&
M_{44}= \kappa \frac{v^2_\eta v_\Sigma}{2v_{\sigma_1}}
+\frac{t_{\sigma_1}}{v_{\sigma_1}},\nonumber \\ \mbox{}&& 
M_{55}=\frac{1}{2\sqrt2}(\lambda_{15}v^2_\chi-\lambda_{16}v^2_\rho)
\frac{v_\eta}{v_{\sigma_2}}+2\lambda_{17}v^2_\eta-\frac{f_2}{4}\frac{v_\rho
v_\chi}{v_{\sigma_2}}\nonumber \\ &+&\frac{t_{\sigma_2}}{v_{\sigma_2}},\quad
M_{66}=\kappa \frac{v^2_\eta v_{\sigma_1}}{2v_\Sigma}+\frac{t_\Sigma}{v_\Sigma} ,
\nonumber \\&&
M_{12}=-\frac{f_1}{2\sqrt2}v_\chi,\;\; M_{13}=
-\frac{f_1}{2\sqrt2}v_\rho,\;\; M_{14}=-\kappa v_\eta v_\Sigma,\nonumber \\
\mbox{}&& 
M_{15}=\frac{1}{2\sqrt{2}} (\lambda_{15} v^2_\chi-
\lambda_{16} v^2_\rho) +2\lambda_{17} v_\eta v_{\sigma_2}, \;\;
M_{16}=\kappa v_\eta v_{\sigma_1},
\nonumber \\&&
M_{23}=-\frac{1}{4}(\sqrt{2}f_1v_\eta+f_2v_{\sigma_2}),
\quad M_{24}=0,\quad M_{25}=\frac{f_2}{4}v_\chi,\nonumber \\ && M_{26}=0,\;\;
M_{34}=0,\quad M_{35}=\frac{f_2}{4}v_\rho,\;\;M_{36}=0, \nonumber \\ &&
M_{45}=0,\;\; M_{46}=-\kappa \frac{v^2_\eta}{2},\;\; M_{56}=0. 
\label{mmi}
\end{eqnarray}

The mass matrix above has two true Goldstone bosons $G^0_{1,2}$ and the
Majoron-like one, $M^0$, and three massive CP-odd pseudoscalar bosons. 
The massless ones are given by
\begin{eqnarray}
G^0_1&=& (0,v_\rho/v_\chi,-1,0,0,0)/(1+v^2_\rho/v^2_\chi)^{1/2} \nonumber 
\\ G^0_2&=& (\frac{v_\eta}{v_{\sigma_2}},-\frac{v_\rho
v^2_\chi}{V_1}, -\frac{v^2_\rho
v_\chi}{V_1},\frac{2v_{\sigma_1}v^2_\Sigma}{V_2},-1,
- \frac{2v^2_{\sigma_1}v_\Sigma}{V_2} )/N \nonumber \\ 
M^0&=&  (0,0,0,
v_{\sigma_1}/v_\Sigma,0,1)/(1+v^2_{\sigma_1}/v^2_\Sigma)^{1/2}
\label{ufa2}
\end{eqnarray} 
where $V_1=v_{\sigma_2}(v^2_\rho+v^2_\chi)$,
$V_2=v_{\sigma_2}(v^2_{\sigma_1}+v^2_\Sigma)$; $N$ is the normalization factor
that we will omit here. We have verify that $M^0$ in Eq.~(\ref{ufa2}) is in fact
the Majoron: by adding an explicit ${\cal F}$-violating term, like $f_3\eta^T
S\eta$, it gets a mass while the other two $G_{1,2}$ remain massless.   
The massive pseudoscalars, for the parameters used before have
the following masses (in GeV): 174, 3625 and 4325. On the other hand, if
$v_{\sigma_1}=0$, which forces $\kappa=0$, the Majoron is purely singlet and the
real and imaginary parts of $\sigma^0_1$ are mass degenerate, i.e., form a
complex field, with mass 
\begin{eqnarray}
m_{\sigma_1}&=& \mu^2_4+\lambda_{10}v^2_{\sigma_2}+
(\lambda_{12}+\lambda_{19})\frac{v^2_\eta}{2}+\lambda_{13}\frac{v^2_\chi}{2}
\nonumber \\ &+&\lambda_{14}\frac{v^2_\rho}{2}+\lambda_{25}\frac{v^2_\Sigma}{2}.
\label{ff}
\end{eqnarray}

We see that in this case the Majoron has not doublet components at all
and it is mainly singlet. Hence it is possible to keep
consistence with LEP data.  
Although there are astrophysical constraints (the Majoron
emission implies a different rate for the stellar cooling) that have to be taken
into account~\cite{astro}, in the basis we
have chosen they are less severe since we have avoided the doublet component of
the Majoron. 
Any way, since these constraints have been already considered in
Ref.~\cite{diaz} and they imply that $v_{\sigma_1}<0.33$ GeV if $v_\Sigma=1$
TeV, we will use these values for  $v_{\sigma_1},v_\Sigma$.

\begin{figure}[ht]
\centering\leavevmode
\epsfxsize=100pt
\epsfbox{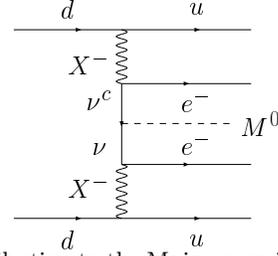}
\vglue -0.01cm
\caption{Contribution to the Majoron emission $(\beta\beta)_{0\nu M}$ decay.
$X^-$ can be a scalar or vector boson.}
\label{fig7}
\end{figure}

\begin{figure}[ht]
\centering\leavevmode
\epsfxsize=100pt
\epsfbox{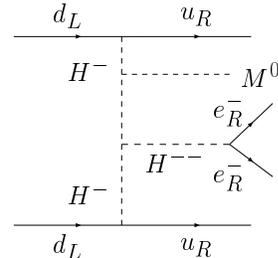}
\vglue -0.01cm
\caption{Trilinear scalar coupling contributing to the 
$(\beta\beta)_{0\nu M}$ decay.}
\label{fig8}
\end{figure}

Once we have shown in what situation there is a safe Majoron-like boson in the
present model we can consider the emission of this Goldstone boson in the
neutrinoless double beta decay. In fact,
as in the triplet Majoron model, in the present model it is possible to have
the neutrinoless double beta decay with Majoron emission: 
$2n\to 2p+2e^-+M^0$~\cite{ggn}, denoted here by $(\beta\beta)_{0\nu M}$.
We will denote the strength of the amplitude of the diagram $i$ of the 
$(\beta\beta)_{0\nu M}$ decay by $B(i)$. 
This decay proceeds via the diagram in Fig.~\ref{fig7} and it has an 
strength proportional to
\begin{equation}
B(7)\propto \frac{m_\nu(v_{\sigma_1}/v_\Sigma)}
{M^4_{X^-}\langle p^2\rangle v_{\sigma_1}},
\label{1m}
\end{equation}
where $X^-$ can be an scalar or a vector boson, {\it i.e.}, the diagram
in Fig.~\ref{fig7} can be formed with anyone of Figs.\ref{fig1}, \ref{fig2} or
\ref{fig3} with a Majoron attached to the neutrinos. The couplings between
neutrinos and the Majoron are diagonal and given by $m_{\nu}/v_{\sigma_1}$. 
Notice that in Eq.~(\ref{1m}) it appears the truly neutrino mass instead of the
effective mass $\langle M_\nu\rangle$ defined in Eq.~(\ref{lifetime}).
However we still can assume that neutrinos have small masses and numerically 
$ m_\nu \approx \langle M_\nu\rangle$.
We will assume also that the contribution to the $(\beta\beta)_{0\nu M}$-decay 
in Fig.~\ref{fig7} with $X^-=W^-$ is the reference one. This diagram  
depends only on the neutrino masses and mixing angles, and we will compare it
with other contributions like the one in Fig.~\ref{fig8}.
The couplings of the Majoron 
to the vector bosons are proportional to $v_{\sigma_1}$ and so they are 
negligible. We will consider only the diagram with the Majoron coupled to the 
scalar $H^-$ since it is proportional to the trilinear $f_2$ shown in 
Eq.~(\ref{3e}). We have
\begin{equation}
B(8)\propto \frac{ff_2}{M^6_{H^-}M^2_{H^{--}}},
\label{1m2}
\end{equation}
with $f$ can be $f_1$ or $f_2$.

Let us consider the ratio 
 \begin{equation}
\frac{B(7)}{A(1)}\propto \frac{m_\nu Q}{32G^2_FM^4_X\langle 
M_\nu\rangle c^4_\theta v_{\Sigma}},
\label{b7a1}
\end{equation}
where we have introduced the factor $Q$ which denotes the available energy. 
It implies that the diagram in Fig.~\ref{fig7} 
is a potentially important contribution when $X$ is the
$W$ vector boson since for $Q\sim 3$ MeV~\cite{jb} the suppression
of $B(7)$ will depend mainly on the value of $v_{\Sigma}$. If $B(7)/A(1)<1$
we obtain that $v_{\Sigma}>1.65\times10^{-2}$ GeV which is automatically
satisfied.
 
\vglue 0.01cm

On the other hand, comparing the amplitudes of the diagrams in Figs.~\ref{fig7} 
and \ref{fig8} we have
\begin{equation}
\frac{B(8)}{B(7)}\propto \frac{ff_2\langle p^2\rangle 
M^4_X v_{\Sigma}}{M^6_{H^-} M^2_{H^{--}}m_\nu},
\label{b8b7}
\end{equation}
and for $M_X=M_V=400$ GeV and $v_{\Sigma}=1$ TeV, using typical values 
as $f=f_1=f_2=-1$ TeV, $M_{H^{-}}=124$ GeV and $M_{H^{--}}=500$ GeV, and 
the other parameters in Eq.~(\ref{b8b7}), we have that $B(8)/B(7)
\approx 1.3\times10^{9}$ or; $B(8)/B(7)\approx 1.5\times10^{3}$ 
if $f_1=f_2=f=-10^{-3}$ TeV.
The relative importance of the processes in Fig.~7 and 8 
will depend on the values of the trilinear parameters and on the
value of $v_{\Sigma}$.

\vglue 0.01cm

\section{Conclusions}
\label{sec:con}

We see that in the 3-3-1 model, like in other models with complicated Higgs 
sector~\cite{sv1,ep}, besides the well known mechanism of exchanging massive
Majorana neutrinos between two standard model $V-A$ vertices, there are
new contributions involving the exchange of scalar
bosons. However, unlike similar mechanism in the context of
extensions of the standard model there is no need of fine tuning in order to
have trilinear scalar couplings giving large contributions to the several
neutrinoless double beta decay modes.
Notice that effective interactions from diagrams like those in Fig.~\ref{fig3}
are still parameterized in the form of two general four-fermion effective 
interactions (they are point-like at 
the Fermi scale) exchanging a light neutrino in between~\cite{pas}. However, 
contributions involving trilinear interactions like those in Figs.~\ref{fig4},
\ref{fig5} and \ref{fig6} necessarily need a six-fermion effective interaction
parameterization.

Another important point to be stressed here is that in the present model
the double Majoron emission: $2d\to 2p+2e^-+2M^0$ may be as contributing as 
the decay with only one Majoron boson. This decay is expected
to be important in supersymmetric models~\cite{mt,pb}. In the present model
it can occur because in diagrams like that in Fig.~\ref{fig8} a second
Majoron can be attached to the scalar lines. Since this coupling is 
proportional to the trilinear $f_{2}$ it is still possible that the 
suppression coming from the mass square in the denominator do not
sufficiently suppresses this process (there is also an important 
contribution coming from the vertex $v_\chi\chi^-\eta^+_1 M^0$). 
There are also contributions 
similar to the one in Fig.~\ref{fig8} but now with the scalar-Majoron vertex 
above being substituted by the vertex $W^-V^+M^0$ which is proportional to 
$g^2v_{\sigma_2}$ and for this reason it is not necessarily suppressed.  
It is interesting to note that this sort of contribution to the
$(\beta\beta)_{0\nu MM}$ decay, coming from adding another trilinear coupling in
the diagram in Fig.~\ref{fig8}, which is not derivatively suppressed, was not
considered in Ref.~\cite{pb}.
Recent experimental data on Majoron emission decays have been constrained
only the effective Majoron-neutrino coupling constant~\cite{jb}.
Other processes like the double $K$ capture~\cite{ggn} can also be important 
in the present model.
 
Some comments are now in order. 
{\it i)} We have not considered possible cancellations 
among several contributions to each diagram. It means that our constraints 
are valid, as we said in Sec.~\ref{sec:bb0n}, for the main component of 
each scalar field of the singly and doubly
charged scalar sectors. 

{\it ii)} Our results were obtained assuming that all new 
contributions, say to the $(\beta\beta)_{0\nu}$ decay, are at most as 
important as the contribution due to a light massive Majorana neutrino 
exchange which is proportional to $\langle M_\nu\rangle$. 
However, we can wonder what would be 
the value of the effective mass $\langle M_\nu\rangle$ if we use the
oscillation data and direct measurements on neutrino masses.
Recent analysis shown that assuming a normal mass hierarchy the effective mass
parameter can take any value from zero to the present upper 
limit~\cite{smirnov}.
In fact, if the data on oscillation is put together with that from
$(\beta\beta)_{0\nu}$ decay and tritium beta decay, it was shown that
if the minimum of $\langle M_\nu\rangle$ with respect to the mixing angles
is greater than the present bound of 0.2 eV, then neutrinos are quasi-Dirac 
particles~\cite{gluza}. As discussed previously, the black box of 
Ref.~\cite{sv1} may induce a negligible Majorana mass to the neutrinos
and in the context of the present model we must interpret this situation 
as an indication of the
fact that the main contribution to the $(\beta\beta)_{0\nu}$ decay is not
the diagram in Fig.~1. In this case the neutrinos would be almost Dirac
particles and the constraints on the several mass scales of the model should
be obtained by comparing directly these contributions with the lower
bound on the half life $T^{0\nu}_{1/2}>1.8\times10^{25}$ yr~\cite{exp}.

{\it iii)} In the basis we have chosen, see Eq.~(\ref{ufa2}), the Majoron
couples at the tree level only to neutrinos, 
and hence the constraints on the $\nu$-$\nu$-$M$ vertex, coming from muon 
($\mu\to e\nu\nu M$), pion and kaon $\pi^+(K^+)\to l\bar{\nu} M$ decays, 
are the same as in Ref.~\cite{pb}. The existence of the
vertex $\nu e^-V^+$, which is proportional to $\sin\theta$ and
for this reason is not relevant for laboratory processes, may have,
as we said before, important astrophysical consequences~\cite{ma1}. 

{\it iv)} phenomenology of non-zero initial electric charge processes, 
like $e^-e^-$ and hadronic ones, will
furnish constraints on the trilinear vertices appearing in Figs.~\ref{fig4},
\ref{fig5} and \ref{fig6} but this will be considered elsewhere. 
The 3-3-1 model has a rich scalar sector indeed. This implies that it
may be, in principle, difficult to separate, in a given process, the
contributions of all fields belonging to a charged sector. However,
it has been shown that in lepton-lepton colliders the left-right asymmetries
are not sensible to the scalar contributions. It means that those
asymmetries are the appropriate observable for the doubly charged
vector bilepton discovery~\cite{assi123}. 
We see that the opposite occurs in the
$(\beta\beta)_{0\nu}$ decay: it is possible that the main contribution
comes from the doubly charged scalar boson, through the diagram in 
Fig.~\ref{fig6}, while the respective vector boson contribution seems to be 
negligible.

Finally we would like to compare our Majoron model with that of Schechter
and Valle~\cite{sv}. Firstly, we notice that although our model has two singlet,
three doublets and a triplet of scalars under the subgroup $SU(2)\otimes U(1)$,
the respective scalar potential is not reduced to the scalar potential 
invariant under the standard $SU(2)\otimes U(1)$ symmetry,
involving the same multiplets. For instance in our model there are cubic
invariants which are not present in the former. Secondly, we have not introduced
right-handed neutrinos and for this reason we have only light neutrinos. It
means that the singlet $\Sigma$ does not couple to the leptons and that the
coupling of neutrinos to Majoron and $Z^0$ are diagonal. Thus, the decays
$\nu_H\to\nu_L +M^0$ and $\nu_H\to \nu_L+\nu^\prime_L+\nu^\prime_L$, are not
induced at the tree level, where $\nu_H$, although light,  is heavier that
$\nu_L$. The decay $\nu_H\to\nu^c_L+M^0$ is produced at the one loop level
due to the mixing between $W$ and $V$. The vertex is proportional to 
$(g^2/\sqrt{2})(v_{\sigma_2}v_{\sigma_1}/v_\Sigma)$; then even
with $v_{\sigma_2}$ of the order of 10 GeV and $v_\Sigma$ of the order of 1 TeV 
the lifetime is of the order of the age of the universe. (The decay
$\nu_H\to\nu_L+M^0$ also occurs but the vertex involved is proportional to
$(g^2/\sqrt{2})(v^2_{\sigma_1}/v_\Sigma)$.)  Notice also that in 
the basis given in Eq.~(\ref{ufa2}) the Majoron does not couple to the charged
leptons so there is not the process $\gamma+e\to M^0+e$ at the tree level which
imposes severe astrophysical constraints in $v_{\sigma_1}$ as has been noted in 
Ref.~\cite{diaz}. 
 
\acknowledgments 
This work was supported by Funda\c{c}\~ao de Amparo \`a Pesquisa
do Estado de S\~ao Paulo (FAPESP), Conselho Nacional de 
Ci\^encia e Tecnologia (CNPq) and by Programa de Apoio a
N\'ucleos de Excel\^encia (PRONEX). 

\appendix

\section{Contraints equation of the scalar potential}
\label{sec:app}
Here we show the constraints equation that must be satisfied by the scalar
potential
\begin{eqnarray*}
&&t_\eta=\mu^2_1v_\eta+\lambda_1v^3_\eta+\frac{\lambda_4}{2}v^2_\rho v_\eta+
\frac{\lambda_5}{2}v^2_\chi v_\eta+\frac{\lambda_{12}}{2}
(v^2_{\sigma_1}+v^2_{\sigma_2})v_\eta \\ &-& 
\frac{\lambda_{15}}{2\sqrt2}v^2_\chi v_{\sigma_2}
+\frac{\lambda_{16}}{2\sqrt2}v^2_\rho v_{\sigma_2}
-\lambda_{17}v^2_{\sigma_2}v_\eta+\frac{\lambda_{19}}{2}v^2_{\sigma_1}v_\eta
\nonumber \\ &-&
\kappa v_\eta v_{\sigma_1}v_\Sigma+\frac{f_1}{2\sqrt2}v_\rho v_\chi, \\ &&
t_\rho=\mu^2_2v_\rho+\lambda_2v^3_\rho+\frac{\lambda_4}{2}v^2_\eta v_\rho+
\frac{\lambda_6}{2}v^2_\chi v_\rho+\frac{\lambda_{14}}{2}(v^2_{\sigma_1}+
v^2_{\sigma_2}) v_\rho\nonumber \\
&+&\frac{\lambda_{16}}{\sqrt2}v_{\sigma_2}v_\eta v_\rho + 
\frac{\lambda_{20}}{4}v^2_{\sigma_2} v_\rho
+\lambda_{23}\frac{v_\rho v^2_\Sigma}{2}\nonumber \\ &+&
\frac{f_1}{2\sqrt2}v_\eta v_\chi 
 +\frac{f_2}{4}v_{\sigma_2}v_\chi , \\ &&
t_\chi=\mu^2_3 v_\chi+\lambda_3v^3_\chi+\frac{\lambda_5}{2}v^2_\eta v_\chi+
\frac{\lambda_6}{2}v^2_\rho
v_\chi+\frac{\lambda_{13}}{2}(v^2_{\sigma_1}+
v^2_{\sigma_2})v_\chi\nonumber \\ &-&
\frac{\lambda_{15}}{\sqrt2}v_{\sigma_2}v_\eta v_\chi +
\frac{\lambda_{18}}{4}v^2_{\sigma_2}v_\chi+\lambda_{24}\frac{v_\chi
v^2_\Sigma}{2} \nonumber \\ &+&
\frac{f_1}{2\sqrt2}v_\eta v_\rho+
\frac{f_2}{4} v_{\sigma_2}v_\rho, \\ &&
t_{\sigma_1}=\mu^2_4
v_{\sigma_1}+\lambda_{10}(v^2_{\sigma_1}+v^2_{\sigma_2})v_{\sigma_1}+
\lambda_{11}v^3_{\sigma_1}+\frac{\lambda_{12}}{2}
v^2_\eta v_{\sigma_1}\nonumber \\ &+&
\frac{\lambda_{13}}{2} v^2_\chi v_{\sigma_1}+
\frac{\lambda_{14}}{2}v^2_\rho v_{\sigma_1}+
\frac{\lambda_{19}}{2}v^2_\eta
v_{\sigma_1}+\frac{\lambda_{25}}{2}v_{\sigma_1}v^2_\Sigma-
\frac{\kappa}{2}v^2_\eta v_\Sigma , \\ &&
t_{\sigma_2}=\mu^2_4
v_{\sigma_2}+\lambda_{10}(v^2_{\sigma_2}+v^2_{\sigma_1})v_{\sigma_2}+
\frac{\lambda_{11}}{2}v^3_{\sigma_2}+\frac{\lambda_{12}}{2}v^2_\eta 
v_{\sigma_2}\nonumber \\ &+&\frac{\lambda_{13}}{2}v^2_\chi v_{\sigma_2}+
\frac{\lambda_{14}}{2}v^2_\rho v_{\sigma_2}
-\frac{\lambda_{15}}{2\sqrt2} v^2_\chi v_\eta
+\frac{\lambda_{16}}{2\sqrt2} v^2_\rho v_\eta 
 - \lambda_{17}v^2_\eta v_{\sigma_2}\nonumber \\ &+&
\frac{\lambda_{18}}{4}v^2_\chi v_{\sigma_2} 
+\frac{\lambda_{20}}{4}v^2_\rho v_{\sigma_2}
+\frac{\lambda_{25}}{2}v_{\sigma_2}v^2_\Sigma 
+\frac{f_2}{4}v_\rho v_\chi, \\ && 
t_{\Sigma}=\mu^2_5 v_\Sigma+\lambda_{21}v^3_\Sigma
+\frac{\lambda{22}}{2}v^2_\eta v_\Sigma+\frac{\lambda_{23}}{2}v^2_\rho v_\Sigma
+\frac{\lambda_{24}}{2}v^2_\chi v_\Sigma \nonumber \\ &+&
\frac{\lambda_{25}}{2}(v^2_{\sigma_1}+v^2_{\sigma_2})v_\Sigma-
\frac{\kappa}{2}v^2_\eta v_{\sigma}.
\end{eqnarray*}


\begin{references}
\bibitem{expsolar}  K. Lande (Homestake 
Collaboration), in {\sl Neutrinos '98}, Proceedings of the XVIII 
International Conference on Neutrino Physics 
and Astrophysics, Japan, 4-9 June 1998, edited by Y. Suzuki and T. Tosuka, to
be published in Nucl. Phys. B (Proc. Suppl.);
Y. Fukuda {\it et al}. (Kamiokande Collaboration), Phys. Rev. Lett. {\bf77},
1683 (1996); T. Kirsten (GALLEX Collaboration) in {\sl Neutrinos '98};
V. Gavrin (SAGE Collaboration) in {\sl Neutrinos '98}; Y. Suzuki 
(SuperKamiokande Collaboration) in {\sl Neutrinos '98}. 
\bibitem{expatm} Y.\ Fukuda {\it et al.} (SuperKamiokande 
Collaboration ), Phys.\ Lett.\ {\bf B433}, 9 (1998); Phys. Rev. Lett. 
{\bf 81}, 1562 (1998); Phys. Lett. {\bf B436}, 33 (1988).
\bibitem{lsnd} C. Athanassopoulos {\it et al.}, 
Phys. Rev. Lett. {\bf77}, 3082 (1996); {\it ibid}, {\bf81}, 1774 (1998).
\bibitem{pdg} D. E. Groom {\it et al.}, The European Physical Journal, 
{\bf C15}, 1 (2000).
\bibitem{rosen} H. Primakov and S. P. Rosen, Rep. Prog. Phys. {\bf22}, 111 
(1969).
\bibitem{rev} For a recent review see H. V. Klapdor-Kleingrothaus, Int. J.
Mod. Phys. {\bf A13}, 3953 (1998); 
P. Fisher, B. Kayser and K. S. McFarland, Ann. Rev. Nucl. Part. Sci. 
{\bf 49} (1999); hep-ph/9906244.
For early references see M. Doi, T. Kotani 
and E. Takasugi, Progr. Theor. Phys. Suppl. {\bf 83}, 1 (1985). 
\bibitem{exp} L. Baudis {\it et al.}, Phys. Rev. Lett. {\bf83}, 41 (1999);
 hep-ex/9902014.
\bibitem{sv1} J. Schechter and J. W. F. Valle, Phys. Rev. D {\bf 25}, 2951
(1982).
\bibitem{pt} V. Pleitez and M. D.  Tonasse, Phys. Rev. D {\bf 48}, 527 (1993).
\bibitem{mv} R. N. Mohapatra and J. D. Vergados, Phys. Rev. Lett. {\bf47}, 
1713 (1981).
\bibitem{hw} W. C. Haxton, S. P. Rosen and G. J. Stephenso, Jr.,
Phys. Rev. D {\bf26} 1805 (1982); 
L. Wolfenstein, {\it ibid}, D {\bf26}, 2507 (1982).
\bibitem{ep} C. O. Escobar and V. Pleitez, Phys. Rev. D {\bf 28}, 1166 (1983).
\bibitem{331} F. Pisano and V. Pleitez, Phys. Rev. D {\bf46}, 410 (1992);
P.\ Frampton, Phys.\ Rev.\ Lett.\ {\bf 69}, 2889 (1992); R.
Foot, O. F. Hernandez, F. Pisano and V. Pleitez, Phys. Rev. D {\bf 47}, 4158
(1993).
\bibitem{srange} M. Hirsch, H. V. Kkapdov-Kleingrothaus and S. G. Kovalenko,
Phys. Rev. Lett. {\bf75}, 17 (1995);
Phys. Rev. D {\bf53}, 1329 (1996); H. V. Klapdor-Kleingrothaus and H. P\"as,
hep-ph/0005045. 
\bibitem{ma1} J. C. Montero, C. A. de S. Pires  and V. Pleitez, Phys. Rev. 
D {\bf 60}, 115003 (1999); hep-ph/9903251.
\bibitem{comment} J. C. Montero, C. A. de S. Pires  and V. Pleitez, 
 Phy. Rev. D {\bf60}, 98701 (1999); hep-ph/9902448.
\bibitem{liung} J. T. Liu and D. Ng, Phys. Rev. D {\bf50}, 548 (1994).
\bibitem{fcnit} M. M. Guzzo et al., hep-ph/9908308. 
\bibitem{joshi} M. B. Tully and G. C. Joshi, Int. J. Mod. Phys. 
{\bf A13}, 5593 (1998). 
\bibitem{sw} S. Weinberg, Phys. Rev. Lett. {\bf29}, 1698 (1972).
\bibitem{gr}  G. B. Gelmini and M. Roncadelli, Phys. Lett. {\bf99B}, 
411 (1981).
\bibitem{concha} M. C. Gonzalez-Garcia and Y. Nir, Phys. Lett. {\bf232}, 383
(1989).
\bibitem{ps} F. Pisano and S. Shelly Sharma, Phys. Rev D {\bf57}, 5670 (1998).
\bibitem{typeII} J. C. Montero, C. A. de Sousa Pires and V. Pleitez,
Phys. Lett. {\bf B502}, 167 (2001); hep-ph/0011296. 
\bibitem{ema} E. Ma, Phys. Rev. Lett. {\bf81}, 1171 (1998).
\bibitem{sv} J. Schechter and J. W. F. Valle, Phys. Rev. D {\bf25}, 774 (1982).
\bibitem{diaz} M. A. Diaz, M. A. Garcia-Jareno, D. A. Restrepo and J. F. W. 
Valle, Nucl. Phys. {\bf B527}, 44 (1998).
\bibitem{astro} D. A. Dicus, E. W. Kolb, V. L. Teplitz and R Wagoner, Phys. 
Rev. D {\bf18}, 1829 (1978). R. N. Mohapatra and P. B. Pal, {\sl Massive
Neutrinos in Physics and 
Astrophysics}, World Scientific, Singapore, 1991; and references therein.
\bibitem{ggn} H. Georgi, S. L. Glashow and S. Nussinov, Nucl. Phys. {\bf B193},
297 (1981).
\bibitem{jb} J. Bockholt {\it et al.}, Phys. Rev. D {\bf54}, 3641 (1996).
\bibitem{pas} H. P\"as, M. Hirsch, H. V. Klapdor-Kleingrothaus and S. G. 
Kovalenko, Phys. Lett. {\bf B453}, 194 (1999).
\bibitem{mt} R. N. Mohapatra and E. Takasugi, Phys. Lett. {\bf 211B}, 192 
(1988).
\bibitem{pb} P. Bamert, C. Burguess and R. N. Mohapatra, Nucl. Phys. 
{\bf B449}, 25 (1995).
\bibitem{smirnov} W. Rodejohann, Phys. Rev. D{\bf62}, 013011 (2000); 
H. V. Klapdor-Kliengrothaus, H. P\"as and A. Yu. Smirnov, Phys. Rev. D {\bf63},
073005 (2001).
\bibitem{gluza} M. Czakon, J. Gluza and M. Zralek, hep-ph/0003161.
\bibitem{assi123} J. C. Montero, V. Pleitez and M. C. Rodriguez, 
Phys. Rev. D {\bf58}, 094026 (1998); Phys. Rev. D {\bf58}, 
097505 (1998); and Int. J. Mod. Phys. {\bf A16}, 1147 (2001). 

\end{references}
\end{document}